\documentstyle[aps,pra,epsfig,twocolumn]{revtex} 

\def\be{\begin{equation}}
\def\ee{\end{equation}}
\def\bea{\begin{eqnarray}}
\def\eea{\end{eqnarray}}
\def\bma{\begin{mathletters}}
\def\ema{\end{mathletters}}

\newcommand{\eins}{\mbox{$1 \hspace{-1.0mm}  {\bf l}$}}
\def\C{\hbox{$\mit I$\kern-.7em$\mit C$}}

\newcommand{\ket}[1]{ | \, #1  \rangle}

\begin{document}
\draft

\title{Equivalence classes of non-local unitary operations}

\author{W. D\"ur$^{1}$ and J. I. Cirac$^{2}$}

\address
{$^1$Sektion Physik, Ludwig-Maximilians-Universit\"at M\"unchen, Theresienstr.\ 37,
D-80333 M\"unchen, Germany\\
$^2$Max Planck Institut f\"ur Quantenoptik, D-85748 M\"uenchen, Germany}

\date{\today}

\maketitle

\begin{abstract}

We study when a multipartite non--local unitary operation can deterministically 
or probabilistically simulate another one when local operations of a certain 
kind ---in some cases including also classical communication--- are allowed. In the 
case of probabilistic simulation and allowing for arbitrary local operations, we 
provide necessary and sufficient conditions for the simulation to be possible. 
Deterministic and probabilistic interconversion under certain kinds of local 
operations are used to define equivalence relations between gates. In the 
probabilistic, bipartite case this induces a finite number of classes.  In 
multiqubit systems, however, two unitary operations typically cannot simulate 
each other with non-zero probability of success. We also show which kind of 
entanglement can be created by a given non--local unitary operation and generalize our 
results to arbitrary operators.

\end{abstract}


\narrowtext

\section{Introduction}

In the last decade, there has been big effort to characterize
qualitatively and quantitatively entanglement properties of pure and
mixed states. This relies in part on the fact that entanglement is
thought to be the key ingredient for many applications in Quantum
Information Theory (QIT). A proper understanding of entanglement is
expected to lead not only to possible new applications in quantum
computation and quantum communication, but also to a more satisfactory
understanding of the basic principles of quantum mechanics and
especially of QIT.

Only quite recently, it was realized that also entanglement properties
of physical operations are of relevance, as after all we deal with {\it
  interactions} in experiments and the {\it
  interactions} allow us to create entangled states. In recent
years, first steps have been taken in this direction. In particular,
the possibility to implement non--local operations consuming an
entangled state \cite{Go98,Ch00,Ei00,Co00,Ci00}, the capability to
create entanglement in an optimal way given an interaction Hamiltonian
\cite{Za00,Du00H,Kr00}, the simulation of an interaction Hamiltonian
by some other one \cite{Du00H,Do01,Ja01,Be01,Le01,Vi01H,Vi01C} as well
as a connection of the entanglement properties of operations to the
entanglement properties of states \cite{Ci00,Du01P} have been
established. Many application of this last relation, including the
storage, tomography, teleportation, cloning and purification of
operations, as well as the possibility to decide whether a given
operation can create entanglement or not, have been found
\cite{Du01P}.

However ---compared to the extensive knowledge on the structure of
entangled quantum states---, only few is known for quantum operations.
The most successful approach to characterize the
entanglement properties of bipartite and multipartite pure states is
concerned with the study of equivalence relations under certain
classes of allowed operations, e.g. local unitaries (LU), local
operations (LO), local operations and classical communication (LOCC)
or stochastic local operations and classical communication (SLOCC),
possible applied to many copies of a system \cite{Be99}.  That is, two
states are identified if they can be obtained from each other by
means of a certain class of operations, e.g. LU.  In this case, we say
that two states belong to the same (equivalence) class under LU. For
bipartite systems, when considering equivalence classes under LU, this
leads to the well known Schmidt decomposition, while the Schmidt
number, i.e.  the number of non--zero Schmidt coefficients, turns out
to be the relevant quantity when considering equivalence classes under
SLOCC \cite{Du00W}. For three qubit systems, this approach allowed to
identify two inequivalent kinds of tripartite entanglement under SLOCC
represented by the states $|GHZ\rangle$ and $|W\rangle$, defined in
Eq. (\ref{WGHZ}) \cite{Du00W}. When applied to many copies of a
bipartite system \cite{Be95}, considering equivalence classes under LOCC
operations and allowing for small imperfections, this criterion leads
to identify all kinds of bipartite pure state entanglement with that
of the EPR--Bohm state $1/\sqrt{2}(\ket{00}+\ket{11})$\cite{Ei35}.

In this paper, we introduce a similar notion of equivalence relations under 
certain classes of operations for {\it non--local unitary operations}. First 
steps in this direction have been recently reported in \cite{Du01C}, where 
equivalence classes for single bipartite unitary operators $U$  under SLOCC were 
considered. That is, whether two non--local unitary operations $U$ and $\tilde 
U$ can simulate each other with non--zero probability of success. In this paper, 
we generalize this approach and provide a general framework which covers also 
single multipartite unitary operations, multiple copies of unitary operations as 
well as other classes of allowed local operations such as LU, LO, and LOCC.  Our 
classification allows us to put a partial order in the set of multipartite nonlocal 
unitary operations.  In the bipartite case of probabilistic simulation, we obtain a complete hierarchic 
classification \cite{Du01C}.

Note that such classification is not only of theoretical interest, but
might also be of some practical relevance. For example, it allows to
decide whether a given unitary operations ---e.g. produced by a weak
interaction--- is already sufficient to implement a relevant
task in QIT, e.g. entanglement purification, one of the basic
primitives for long range quantum communication using quantum
repeaters \cite{Br98}. Standard entanglement purification schemes \cite{Be96} require the
possibility to implement CNOT gates \cite{noteCNOT,noteGate}. In certain
physical systems ---e.g. the polarization modes of singe photons---,
one might however not be able to implement a CNOT gate but rather only
a weak interaction between two particles, e.g. a phase gate with a
small phase. It is a relevant question in this context whether such
gates are already sufficient to implement a CNOT gate ---and thus
entanglement purification--- with certain probability of success. We
derive necessary and sufficient conditions in terms of SLOCC
equivalence of pure states.  This allows e.g. to answer the problem
mentioned above in a positive way, i.e. any phase gate with an
arbitrary small phase allows to realize entanglement purification.
Note that in the context of entanglement purification, it is
sufficient to consider only probabilistic simulation of the gates, as
entanglement purification itself is already a probabilistic process.

In the context of quantum computation, however, such probabilistic
simulation may change the complexity class of a given algorithm and
might thus not be suitable. In this case, deterministic simulation
plays a more important role, which corresponds to equivalence classes
of unitary operations under deterministic LOCC.


The paper is organized as follows: We start in Sec. \ref{Review} by
reviewing some relevant results on equivalence classes under LU, LOCC
and SLOCC for pure states and a previously introduced isomorphism
which relates non--local physical operations and states.  In Sec. \ref{Basics},
we fix some notation and define in a similar way gate simulation and
equivalence classes under LU, LOCC and SLOCC for {\it unitary
  operations}. We also  briefly discuss gate simulation under LU and LOCC in 
Sec. \ref{LULOCC}. The isomorphism of Sec. \ref{Review} turns out to be
the main tool to establish necessary and sufficient conditions for
probabilistic gate simulation, as it provides a connection of this
problem to the well studied problem of SLOCC conversion of pure
states. In Sec. \ref{SLOCCclasses}, this connection is established and
the implications for bipartite and multipartite unitary operations are
discussed in detail. Sec. \ref{create} is concerned with the question,
which kind of entanglement a given unitary operation can create, and
necessary and sufficient conditions in terms of LU, LO, LOCC and SLOCC conversion of
pure states are given. In Sec.  \ref{General}, we generalize our
results to arbitrary operators and summarize and conclude in
Sec. \ref{Summary}.



\section{Relevant results for pure states and connection between states and operations}\label{Review}

In this section, we review some relevant results on equivalence
classes under certain classes of operations for multipartite pure
states as well as the isomorphism between physical
operations and states.

\subsection{Equivalence classes under LU, LOCC and SLOCC for pure states}\label{SLOCCpure}

A widely studied subject in QIT is concerned with the entanglement
properties of multiparticle pure states. A situation of particular
interest consists of several spatially separated parties, each of them
holding one of the systems of a multiparticle pure state they share.
In this setting, the parties are restricted to apply some kind of {\it
  local} operations and eventually to communicate classically. In this
scenario, it turned out to be a very fruitful approach to identify
pure states which can be converted into each other using a certain
kind X of local operations, where X$\in\{$LU, LU+ancilla, LO,  LOCC,
SLOCC$\}$ \footnote{One may also include in this list catalytic
  assisted operations of each kind.} \cite{Be99}, as it allows to
identify in some sense the entanglement properties of two states. In
the case of LU, LU+ancilla, LO and LOCC, both states allow to perform
exactly the same QIT tasks, while in case of SLOCC operations, the
probability of a successful performance of the task may differ.

We say that a multipartite entangled pure state $|\Psi\rangle$ can be converted 
to some other pure state $|\Phi\rangle$ under a certain class of operations X, 
$|\Psi\rangle \rightharpoonup_{\rm X} |\Phi\rangle$, if there exists a sequence 
of local operations of the kind X which transform the state $|\Psi\rangle$ 
exactly into the state $|\Phi\rangle$. In case of SLOCC, only a 
probabilistic conversion is required, i.e. the state $|\Phi\rangle$ has to be 
obtained only with some non--vanishing probability of success.

Note that above relation induces an equivalence relation in the set of pure 
states, namely two states $|\Psi\rangle$ and $|\Phi\rangle$ are equivalent under 
operations of the kind X, $|\Psi\rangle \rightleftharpoons_{\rm X} 
|\Phi\rangle \Leftrightarrow |\Psi\rangle \rightharpoonup_{\rm X} |\Phi\rangle$ and 
$|\Psi\rangle \leftharpoondown_{\rm X} |\Phi\rangle$. That is, the states can be 
converted into each other and are said to belong to the same equivalence class. 

\subsubsection{Bipartite systems}

The definition of equivalence classes under LU allows to write any
pure state $|\Psi\rangle\in\C^d\otimes\C^d$ in its Schmidt
decomposition, i.e. to identify $|\Psi\rangle$  with a state
of the normal form
\be
\sum_{i=1}^{n_{\Psi}} \sqrt{\lambda_i} \ket{i}\otimes\ket{i} =
U_A\otimes U_B\ket{\Psi}; ~~~ n_{\Psi} \leq d,\label{Schmidt1},
\ee
where $S=\{|i\rangle\}_{i=1}^d$ is an orthonormal basis. The real,
positive coefficients $\lambda_i\not =0$ sum up to unity and we have that
$n_{\Psi}$ is the so called Schmidt number, i.e. the number of nonzero
Schmidt coefficients. The Schmidt decomposition turned out to be a
very useful tool in many applications in QIT.

It is known that considering equivalence classes under LOCC does not
allow for a further reduction of the relevant parameters and thus for
no further simplification. This is due to the fact that two pure states
$\ket{\Psi}$ and $\ket{\Phi}$ can be obtained with certainty from each
other by means of LOCC if and only if they are related by LU
\cite{Vi00J,Be99}. Thus one has to deal with infinitely many classes
(i.e. kinds of entanglement) even in the simplest scenario of bipartite
two level systems \cite{Li97,Su00,Ca00,Ke99}.

Considering equivalence classes under SLOCC, however, allows for a
further simplification and to identify a finite number of $d$
inequivalent classes. As shown e.g.  in Ref. \cite{Du00W}, two
entangled pure states $|\Psi\rangle$ and $|\Phi\rangle$ are equivalent
under SLOCC, $|\Psi\rangle\rightleftharpoons_{\rm SLOCC} |\Phi\rangle$
$\Leftrightarrow$ $n_\Psi=n_\Phi$, i.e. they have the same Schmidt number. Conversion of
$|\Psi\rangle$ to $|\Phi\rangle$ under SLOCC, $|\Psi\rangle
\rightharpoonup_{\rm SLOCC} |\Phi\rangle$ is possible if and only if $n_\Psi \ 
\geq n_\Phi$. This provides a complete, hierarchic classification for
bipartite pure states of arbitrary dimension.

The concept of equivalence classes can be applied also to many copies
of a bipartite system. In this case, it turns out \cite{Be95} that
equivalence under deterministic LOCC ---when allowing for small
imperfections--- leads to identifying all
bipartite pure-state entanglement with that of the EPR-Bohm state
$1/\sqrt{2}(\ket{00}+\ket{11}) $\cite{Ei35}. That is, the entanglement
of any pure state $\ket{\psi}_{AB}$ is asymptotically equivalent,
under deterministic LOCC, to that of the EPR-Bohm state, the entropy of
entanglement $E(\psi_{AB})$ ---the entropy of the reduced density
matrix of either system $A$ or $B$--- quantifying the amount of EPR-Bohm
entanglement contained asymptotically in $\ket{\psi}_{AB}$.

\subsubsection{Multipartite systems}\label{SLOCCpuremulti}

In multipartite systems, equivalence classes under LU have also been
studied, however no such particularly simple form as the Schmidt
decomposition for bipartite systems could be obtained (see however
Ref. \cite{Ca00,Ac99}). 
This lack of Schmidt decomposition is one of the reasons for our still 
restricted knowledge on the entanglement properties of multipartite
pure states. 

When considering equivalence classes under SLOCC in three qubit
systems, this classification
allowed to identify two inequivalent kinds of true tripartite
entanglement \cite{Du00W}, represented by the states
\bma\bea\label{WGHZ}
|W\rangle&=&1/\sqrt{3}(|001\rangle+|010\rangle+|100\rangle)\label{W},\\
|GHZ\rangle&=&1/\sqrt{2}(|000\rangle+|111\rangle),
\eea\ema
respectively. The corresponding equivalence classes under SLOCC are
called W--class and GHZ--class. In multipartite systems
or tripartite systems of higher dimension, infinitely many classes
under SLOCC exist, so typically two multipartite entangled states
cannot be transformed into each other with non--zero probability of
success \cite{Du00W}. Remarkably, all equivalence classes for four
qubit systems have been identified recently \cite{Ver01}.

\subsection{Isomorphism between physical operations and states}\label{Isomo}

In \cite{Ci00}, an isomorphism which relates non--local physical operations
[equivalently completely positive maps (CPM) ${\cal E}$] acting on two
systems and (unnormalized) states (positive operators $E$) was
introduced and generalized to $N$--partite systems in \cite{Du01P}.
When applied to unitary operations $U$, it turns out that the
corresponding state is pure. We review this isomorphism ---specialized
to unitary operations--- in detail, as it provides the proper tool to
connect the problem of classification of operations under SLOCC to the
well studied problem of classification of pure states under SLOCC.

We consider several spatially separated systems $A,B, \ldots ,Z$, each
possessing several $d$--level systems.  
Let
\be
\label{MES} |\Phi\rangle_{A_{1,2}} =
\frac{1}{\sqrt{d}}\sum_{i=1}^d |i\rangle_{A_1}\otimes
|i\rangle_{A_2},
\ee
be a maximally entangled state (MES), $|\Phi\rangle \in \C^{d^2}$.  We
denote by $P_\Phi^{A_{1,2}} \equiv |\Phi\rangle_{A_{1,2}}\langle\Phi|$
a projector on this state. 

We consider a $N$--partite unitary operation $U$ acting on several
$d$--level systems, one located in each site $A, B,\ldots ,Z$. Let
$|\Psi_U\rangle_{A_{1,2}\ldots Z_{1,2}} \in (\C^{d^2})^{\otimes N}$ be
a $N$--partite pure state and $P_{\Psi_U}\equiv
|\Psi_U\rangle\langle\Psi_U|$ the corresponding projector on this
state. As shown in \cite{Ci00,Du01P}, one obtains the following
relations between the unitary operation $U$ and a pure state
$|\Psi_U\rangle$:
\bma\label{iso}\bea
|\Psi_U\rangle_{A_{1,2}\ldots Z_{1,2}} &=& U_{A_1\ldots Z_1} (|\Phi\rangle_{A_{1,2}}\otimes \ldots \otimes |\Phi\rangle_{Z_{1,2}}),\label{iso1}\\
U\rho_{A_1\ldots Z_1}U^{\dagger}&=& d^{2N}{\rm tr}_{A_{2,3}\ldots
Z_{2,3}} \nonumber\\ &&(P_{\Psi_U}^{A_{1,2} \ldots Z_{1,2}} \rho_{A_3\ldots
Z_3} P_{\Phi}^{A_{2,3}}\ldots P_{\Phi}^{Z_{2,3}})\label{iso2}.
\eea\ema
These equations have a very simple interpretation: On one hand,
(\ref{iso1}) states that $|\Psi_U\rangle$ can be created from a
$N$--party product state deterministically, given a single application
of the unitary operation $U$, where each party prepares locally a MES.
On the other hand, (\ref{iso2}) tells us that given $|\Psi_U\rangle$
(particles $A_{1,2}B_{1,2}\ldots Z_{1,2})$, one can implement the
multi---particle operation $U$ on an arbitrary state $\rho$ of $N$
$d$--level systems (particles $A_3B_3\ldots Z_3$) probabilistically, by
measuring locally the projector $P_\Phi$ on particles
$A_{2,3},B_{2,3},\ldots ,Z_{2,3}$ in each of the locations. Note that
the probability of success is given by $p=1/d^{2N}$.

Exactly the same relations hold when considering an arbitrary 
operator $O$ instead of the unitary operation $U$. Also in this case,
the corresponding state, $|\Psi_O\rangle$, is pure. We have that to
each $N$--partite operator $U$ [$O$] corresponds one unique pure state
in $(\C^{d^2})^{\otimes N}$, and to any such pure state corresponds an
 operator O. However, not any such pure state corresponds to a
{\it unitary} operation $U$.


\section{Definition of equivalence classes for non--local unitaries}\label{Basics}

Motivated by the various insights following from the definition of
equivalence classes under certain classes of operations for pure
states (see Sec. \ref{SLOCCpure} for details), we define a similar
notion for unitary operations.  We consider $N$ spatially separated,
$d$--dimensional systems and non--local unitary operations $U,{\tilde
  U}\in SU(d^N)$ acting on those systems.  We allow for a certain
restricted class of {\it local} operations, e.g. LU, and are
interested in the simulation of non--local unitary operations under
these conditions.

{\bf Definition 1:} A unitary operation $U$ can simulate $\tilde U$
under a specific class of operations $X$, $U \rightharpoonup_{\rm X}
\tilde U$, where $X\in\{$LU, LU+{\rm ancilla}, LO, LOCC, SLOCC$\}$ if the
action of $\tilde U$ on any input state $\rho$ can be obtained using a
sequence of operations of the kind X applied before and after a single
application of $U$.

Note that in the case of LU, LU+ancilla, LO and LOCC operations, the
simulation has to be deterministic, while for SLOCC operations, only a
certain non--zero probability of success is required. In the case of
LU+ancilla, LO, LOCC and SLOCC operations, also additional auxillary systems
are allowed and it is not required that the operation $U$ has to be
performed on the input state $\rho$ directly. We only demand that the
total action of the sequence of operations we perform ---after tracing
out auxillary systems--- is given by $\tilde U\rho\tilde U^{\dagger}$ for 
arbitrary input states $\rho$.

{\bf Definition 2:} Two unitary operations $U$ and $\tilde U$ are
equivalent under operations of the class X, $U \rightleftharpoons_{\rm
  X} \tilde U$ if $U \rightharpoonup_{\rm X} \tilde U$ and $U
\leftharpoondown_{\rm X} \tilde U$, i.e. the two operations can
simulate each other under the class of operations X. 

Definition 2 defines an equivalence relation in the set of non--local unitary 
operations and thus allows to identify equivalence classes. Two non--local operations 
$U, \tilde U$ belong to the same equivalence class under operations of the kind 
X  if $U \rightleftharpoons_{\rm X} \tilde U$. Together with definition 1, this 
allows to obtain a partial order in the set of unitary operations. The aim of 
this paper is to identify equivalence classes for non--local unitary operations 
under a specific class of operations. In what follows, we will mainly focus on 
equivalence classes under SLOCC. 

Note that the same definitions also makes sense when applied to
multiple copies of operations, i.e. $U=V^{\otimes N}$ and $\tilde
U=\tilde V^{\otimes M}$, where $N$ and $M$ may be different. In this
case, one is concerned with the question whether $N$ simultaneous
applications of a given operation $V$ can simulate $M$ simultaneous
applications of an operation $\tilde V$.

We would also like to point out that the problem of simulation of unitary 
operations is not equivalent to the problem of Hamiltonian simulation. In the 
latter case, intermediate local operations can be applied, while in the former 
case, one considers a fixed, non--local unitary operation $U$ ---e.g. given by 
some black box---, and local operations can only be applied before and after the 
application of $U$. That is, the process of interaction is inaccessible for some 
reason, e.g. because it is taking place at a very short timescale.


\section{Equivalence classes under LU and LOCC for non--local unitary operations}\label{LULOCC}

In this section, we review some results on equivalence classes under
LU and LOCC for unitary operations. We show that for unitary
operations ---in contrast to pure state conversion---, equivalence
under LU is not the same as equivalence under LOCC.

\subsection{Equivalence classes under LU}

The only well studied example of equivalence classes under LU for
unitary operations is the case of two qubits, i.e. $U \in SU(2^2)$
\cite{Kr00}. Kraus et. al showed in \cite{Kr00} (see also \cite{Kh01}) that any bipartite unitary
operations $U$ acting on two qubits can uniquely \cite{noteunique} be written as  
\bma\label{Ugen}\bea
U_{AB}&=&V_A\otimes W_B e^{-iH} \tilde V_A\otimes \tilde
W_B,\\
H&=&\sum_{i=1}^3H_i, ~~~H_i=\mu_i \sigma_i^A\otimes \sigma_i^B, \\
&&\pi/4\geq \mu_1\geq\mu_2 \geq |\mu_3| \geq 0.
\eea\ema
That is, any unitary operation $U$ can up to local unitaries be
written in the normal form $e^{-iH}$, which might be compared to the
Schmidt decomposition for bipartite pure states. Two unitary operations
$U$ and $\tilde U$ belong to the same equivalence class under LU if and only if 
$\mu_i(U) = \mu_i(\tilde U) \forall i$.  This normal form already
turned out to be useful in a number of applications \cite{Kr00,Du01P}.
No similar result is known for unitary operations acting on higher
dimensional systems or multipartite operations.

\subsection{Equivalence classes under LOCC}

Only few is known also on deterministic simulation of unitary operations under 
LOCC. Very recently, deterministic simulation of two--qubit unitary operations 
given a CNOT [SWAP] operation \cite{noteCNOT} were studied in Ref. \cite{Du01C}. In 
Ref. \cite{Vi01C}, it was shown that catalytic equivalence under LOCC ---i.e. 
allowing to use in addition to LOCC some entangled state, which has to be given 
back undisturbed at the end of the process--- is not the same as equivalence 
under LOCC. In fact, deterministic gate simulation under catalytic LOCC turns 
out to be possible in some cases where it is impossible under LOCC alone. Also 
equivalence under LU and equivalence under LU+ancilla turned out to be different 
\cite{Vi01H}.

We show that in contrast to what happens for pure state conversion,
equivalence under LU(+ancilla) and equivalence under LOCC are
different when considering unitary operations.  To see this, we
consider two copies of the CNOT operation, $U_{\rm
  CNOT}^{\otimes 2}$, and the SWAP operation, $U_{\rm
  SWAP}$ \cite{noteCNOT}. It is easy to show that $U_{\rm CNOT}^{\otimes 2}
\rightleftharpoons_{\rm LOCC} U_{\rm SWAP}$, while $U_{\rm
  CNOT}^{\otimes 2} \not\rightleftharpoons_{\rm LU} U_{\rm SWAP}$. The
first relation can be checked by noting that the CNOT operation can
create 1 ebit of entanglement, while the SWAP operation can create 2
ebits out of a product state. Given the facts that (i) using classical
communication and one ebit of entanglement, one can implement a CNOT
gate, and (ii) given 2 ebits of entanglement plus classical communication,
one can implement a SWAP operation (see e.g. \cite{Co00}), it readily follows that the two
operations in question can simulate each other under LOCC.

On the other hand, one can show that 
$U_{\rm CNOT}^{\otimes 2} 
\not\leftharpoondown_{LU} U_{\rm SWAP}$ \cite{Vi02}. The impossibility of 
this 
process is based on the fact that the classical communication capacity of 
two CNOT operations is given by 2 bits, while the classical communication 
capacity of the SWAP operation is just one bit. This last property follows from 
the fact that whenever one applies a SWAP operation to an arbitrary (possibly 
locally manipulated) product input state, a two--dimensional subspace coming 
from Alice side appears at Bobs (and vice versa). However, a two dimensional 
subspace cannot contain more than one bit of classical information. Since for 
complete simulation of operations under LU, it is required that exactly the same 
tasks can be performed (including also classical information transmission, which 
has to be considered as a resource in this scenario) and due to the fact that LU 
do not change the classical capacity, this implies that the two operations in 
question are not equivalent under LU.



\section{Equivalence classes under SLOCC for non-local 
  unitaries}\label{SLOCCclasses}

In this section, we establish a connection between equivalence classes 
under SLOCC for unitary operations as stated in Sec. \ref{Basics} and
equivalence classes under SLOCC for entangled pure states (see Sec.\ref{SLOCCpure}).
The isomorphism (\ref{iso}), discussed in detail in Sec. \ref{Isomo},
turns out to be the central tool. This relation is expressed in the
following   

{\bf Result 1}: $U$ can simulate $\tilde{U}$ under SLOCC ($U
\rightharpoonup_{\rm SLOCC} \tilde U$) $\Longleftrightarrow$ 
$|\Psi_U\rangle$ can be converted to $|\Psi_{\tilde{U}}\rangle$ by means of SLOCC
($|\Psi_U\rangle \rightharpoonup_{\rm SLOCC} |\Psi_{\tilde{U}}\rangle)$.

{\it Proof:}
$(\Rightarrow):$ Given that $U$ can simulate ${\tilde U}$, it is easy
to show that $|\Psi_U\rangle$ can be converted to
$|\Psi_{\tilde{U}}\rangle$ by means of SLOCC. The conversion takes
place as follows: According to (\ref{iso2}),
$|\Psi_{\tilde{U}}\rangle$ can be used to implement $U$ with certain
probability of success. Now a single application of $U$ allows to
simulate ${\tilde U}$ probabilistically. According to (\ref{iso1}),
${\tilde U}$ can be used to create $|\Psi_{\tilde{U}}\rangle$ out of a
product state, which finishes the proof in
one direction.\\
$(\Leftarrow):$ Given that $|\Psi_U\rangle$ can be converted to
$|\Psi_{\tilde{U}}\rangle$ by means of SLOCC, we have to show that $U$
can simulate $\tilde{U}$ probabilistically. The proof goes as follows:
$U$ is used to create the state $|\Psi_U\rangle$ using (\ref{iso1}),
which can be converted by means of SLOCC to
$|\Psi_{\tilde{U}}\rangle$. Now $|\Psi_{\tilde{U}}\rangle$ can be used
to implement ${\tilde U}$ with certain probability of success
according to (\ref{iso2}), which finishes the proof of the statement.

Note that from Result 1 follows that two unitary operations $U$ and
$\tilde U$ belong to the same equivalence class under SLOCC if and only if the corresponding
pure states $|\Psi_U\rangle$ and $|\Psi_{\tilde{U}}\rangle$ are
equivalent under SLOCC, i.e. they can be
converted into each other by means of SLOCC \cite{Du00W},
\be
U\rightleftharpoons_{\rm SLOCC} \tilde U \mbox{    }
\Longleftrightarrow 
|\Psi_U\rangle \rightleftharpoons_{\rm SLOCC} |\Psi_{\tilde{U}}\rangle\label{EquivU}.
\ee
As LOCC are included in SLOCC, it also follows that equivalence of
unitary operations under SLOCC is a necessary (but not sufficient)
condition for their equivalence under LOCC.

In the following, we are going to illustrate Result 1 and Eq.
(\ref{EquivU}) and apply it to bipartite and multipartite unitary
operations.

\subsection{Bipartite unitary operations}

In this section, we apply Result 1 to bipartite unitary operations. The results 
we derive here are already obtained in Ref. \cite{Du01C}, however we review the 
derivation in detail in order to illustrate Result 1. We are going to show 
that there always exists a finite set of inequivalent classes under SLOCC of 
bipartite unitary operations which are hierarchically ordered. For unitary 
operations $U\in SU(d^2)$ acting on two $d$--level systems, at most $d^2$ 
inequivalent classes exist. In the case of two quibit unitary operations, i.e. 
$d=2$, only three classes remain. The operations CNOT and SWAP \cite{noteCNOT} 
appear as natural representatives.

From Sec. \ref{SLOCCpure} and Result 1 follows that the relevant
quantity that determines the equivalence class under SLOCC for a unitary
operation $U$ is the Schmidt number $n_{\Psi_U}$ of the corresponding
pure state $|\Psi_U\rangle$. This follows from the fact that two
bipartite pure states are equivalent under SLOCC if and only if they have the
same Schmidt number (see Sec. \ref{SLOCCpure}). That is, two unitary operations $U$ and $\tilde U$ are
equivalent under SLOCC if and only if the corresponding states $|\Psi_U\rangle,
|\Psi_{\tilde U}\rangle$ (see Eq.(\ref{iso1})) have the same Schmidt
number, i.e. $n_{\Psi_U} = n_{\Psi_{\tilde U}}$. A unitary operation $U$ can
simulate another operation $\tilde U$ under SLOCC $\Leftrightarrow$ $n_{\Psi_U} \geq
n_{\Psi_{\tilde U}}$, which provides the announced hierarchic
classification. Since $n_{\Psi_U} \leq d^2$ for $U\in SU(d^2)$, one obtains the announced upper bound for the number of classes, $d^2$.

When applied to unitary operations $U\in SU(2^2)$ acting on two
qubits, one finds that at most four inequivalent classes exist,
corresponding to Schmidt numbers $n_{\Psi_U}\in \{1,2,3,4\}$. We will
show, however, that the case $n_{\Psi_U}=3$ does not exist. That is,
the pure state corresponding to any bipartite unitary operation $U$
has either Schmidt number $n_{\Psi_U}$ 1,2 or 4. Recall that although
to any unitary operation corresponds a unique pure state via Eq.
(\ref{iso1}), not to any pure state corresponds a unitary operation.
In fact, it turns out that all pure states with Schmidt number 3 do not
correspond to a unitary operation (but to some non--unitary
operator $O$).

To see this, recall that any bipartite unitary operation acting on two 
qubits can be written as indicated in Eq. (\ref{Ugen}).
Using that $\eins_{A_1}\otimes V_{A_2} |\Phi\rangle = V^T_{A_1}\otimes
\eins_{A_2} |\Phi\rangle$, where $|\Phi\rangle$ is defined in Eq.
(\ref{MES}) and $d=2$, we can write the state $|\Psi_U\rangle$
corresponding to $U$ via Eq. (\ref{iso1}) as
\be
|\Psi_U\rangle=V_{A_1}\otimes W_{B_1}\otimes \tilde V^T_{A_2} \otimes 
\tilde W^T_{B_2}e^{-iH_{A_1B_1}}|\Phi\rangle_{A_{1,2}}|\Phi\rangle_{B_{1,2}}.
\ee
The local unitary operations $V,W,\tilde V,\tilde W$ do not
change the Schmidt number of $|\Psi_U\rangle$, so we may set them to
$\eins$ without loss of generality. We denote a orthogonal basis of 
maximally entangled states by $\{\Phi_i\rangle\}_{i=0,1,2,3}$ with 
\be 
|\Phi_i\rangle\equiv\sigma_i\otimes \eins |\Phi\rangle,\label{Bell}
\ee
and introduce the shorthand notation $c_{\mu_i}\equiv \cos(\mu_i)$, $s_{\mu_i} \equiv
\sin(\mu_i)$. Using that
$e^{-i(H_1+H_2+H_3)}=e^{-iH_1}e^{-iH_2}e^{-iH_3}$, were $H_i$ are
defined in Eq. (\ref{Ugen}), we find that 
\be
|\Psi_U\rangle= \sum_{k=0}^{3} a_i |\Phi_i\rangle_{A_{1,2}}|\Phi_i\rangle_{B_{1,2}},
\ee   
which already corresponds ---up to some irrelevant phase factors--- to the
Schmidt decomposition. The coefficients $a_i$ are given by 
\bma\bea
a_0&=&c_{\mu_1}c_{\mu_2}c_{\mu_3}-is_{\mu_1}s_{\mu_2}s_{\mu_3},\\ 
 a_1&=&c_{\mu_1}s_{\mu_2}s_{\mu_3}-is_{\mu_1}c_{\mu_2}c_{\mu_3},\\ 
 a_2&=&s_{\mu_1}c_{\mu_2}s_{\mu_3}-ic_{\mu_1}s_{\mu_2}c_{\mu_3},\\ 
 a_3&=&s_{\mu_1}s_{\mu_2}c_{\mu_3}-ic_{\mu_1}c_{\mu_2}s_{\mu_3}. 
\eea\ema
It is now straightforward to check that whenever one demands that one of
the coefficients $a_i$ should be zero (which corresponds to having a
Schmidt number 3 or less), automatically also a second coefficient (or 
even two others)
vanishes. E.g., $a_0=0$ implies that
$c_{\mu_1}c_{\mu_2}c_{\mu_3}=s_{\mu_1}s_{\mu_2}s_{\mu_3}=0$. Assuming
e.g. that $c_{\mu_1}=s_{\mu_2}=0$, one finds that also $a_3=0$ etc..  
This implies that $|\Psi_U\rangle$ cannot have $n_{\Psi_U}=3$ and thus 
no bipartite unitary operation with this property exists.

Thus there exist three classes of $SU(4)$ unitary operations under SLOCC:

\begin{itemize}
  
\item{Class 1: $n_{\Psi_U}=1:$} This are {\it local} unitary
  operations, with $\mu_1=\mu_2=\mu_3=0$ in Eq. (\ref{Ugen}).
  
\item{Class 2: $n_{\Psi_U}=2:$} This are nonlocal unitary operations
  with $\mu_1\not=0$ and $\mu_2=\mu_3=0$ in Eq. (\ref{Ugen}). It is
  natural to choose the corresponding state to be a maximally
  entangled state with Schmidt number $n_{\Psi_U}=2$ as a
  representative in this case, which leads to the CNOT operation
  \cite{noteCNOT} as a natural representative of this class. Note that
  the CNOT is up to local unitaries equivalent to an operation of the
  form (\ref{Ugen}) with $\mu_1=\pi/4,\mu_2=\mu_3=0$. This can be seen by noting
  that \be |\Psi_{U_{\rm
      CNOT}}\rangle=\frac{1}{\sqrt{2}}(|00\rangle_{A_{1,2}}|\Phi_0\rangle_{B_{1,2}}+|11\rangle_{A_{1,2}}|\Phi_1\rangle_{B_{1,2}}).
  \ee 
  
\item{Class 3: $n_{\Psi_U}=4:$} This are nonlocal unitary operations
  with $\mu_1, \mu_2\not=0$ and $\mu_3$ arbitrary. One may choose as a
  representative an operation which corresponding state is a maximally
  entangled state with Schmidt number $n_{\Psi_U}=4$. This leads to
  the SWAP operation \cite{noteCNOT} as a natural representative of
  this class. An operation of the form (\ref{Ugen}) with
  $\mu_1=\mu_2=\mu_3=\pi/4$ is up to local unitaries equivalent to the
  SWAP operation. This can be seen by noting that \be |\Psi_{U_{\rm
      SWAP}}\rangle=|\Phi_0\rangle_{A_1B_2}|\Phi_0\rangle_{A_2B_1}.
  \ee 

\end{itemize} 
Recall that any operations of class 3 can simulate operations of class
2 under SLOCC, however the reverse process is not possible. This
implies on the one hand that any non--local unitary operation can be
used to simulate a CNOT operation probabilistically (and thus to
implement entanglement purification), while the CNOT operation can
e.g. not be used to simulate $e^{-it(\sigma_x^A\otimes\sigma_x^B +
  \sigma_y^A\otimes\sigma_y^B)}$ with non--zero probability of success
even for $t\ll 1$.
The procedure sketched in the proof of Result 1 also provides a
practical protocol to achieve this task.

\subsection{Multipartite unitary operations}

For multipartite unitary operations $U\in SU(d^N)$ acting on $N$
$d$--level systems, $N\geq 3$, we have that the corresponding pure
state $|\Psi_U\rangle\in (\C^{d^2})^{\otimes N}$. As even for $d=2$,
typically two entangled pure states of this kind are not equivalent
under SLOCC (This follows from the fact that in multipartite systems
---except the case of three qubits--- infinitely many equivalence
classes exist \cite{Du00W}), Result 1 leads us to expect that
typically two multipartite unitary operations $U$ and $\tilde U$ will
be inequivalent under SLOCC. We can not offer a formal proof of this
statement, because equivalence classes under SLOCC for multipartite
pure states have not been completely identified yet (important
exceptions are all bipartite systems and systems of three and four
qubits). However, we illustrate with the help of a simple example that in
fact infinitely many inequivalent classes under SLOCC of unitary
operations exist in the case of fourpartite unitary operations acting
on qubits.

To this aim, we consider a one parameter family of unitary
operations $U(t)\in SU(2^4)$, generated by the interaction Hamiltonian
$H$ applied for some time $t$, $0<t<\pi/4$, 
i.e. $U(t)\equiv e^{-iHt}$, with
\bea
H&=&\sigma_x^A\otimes\sigma_x^B\otimes\sigma_x^C\otimes\sigma_x^D+\eins^A\otimes\eins^B\otimes\sigma_x^C\otimes\sigma_x^D\nonumber\\
&&+\sigma_x^A\otimes\sigma_x^B\otimes\eins^C\otimes\eins^D.
\eea
We show that $U(t)$ and $U(\tilde t)$ are inequivalent under SLOCC if $t
\not=\tilde t$. This might be surprising at the first sight, because
this means that a unitary operation generated by a certain interaction
switched on for a certain time $t$ cannot be used to simulate ---not
even probabilistically--- a unitary evolution generated by the {\it
  same} interaction, switched on e.g. for some smaller time $\tilde
t$. This is in contrast to what happens for bipartite unitary
operations.

It is easy to show that $U(t)$ is of the form
$U(t)=\alpha(t)\eins+\beta(t)H$ with complex coefficients
$\alpha(t),\beta(t)$. This implies that the state
$|\Psi_{U(t)}\rangle$ corresponding to $U(t)$ via (\ref{iso1})
---after a local basis change $|\Phi_0\rangle \rightarrow |0\rangle$
and $|\Phi_1\rangle \rightarrow |1\rangle$ in all four locations---
can be written as
\be
|\Psi_{U(t)}\rangle=\alpha(t)|0000\rangle+\beta(t)(|1111\rangle+|0011\rangle+|1100\rangle).
\ee
It is now straightforward to check ---applying the results of
Verstraete {\it et.al} \cite{Ver01}--- that if $t \not= \tilde t$, then
$|\Psi_{U(t)}\rangle \not\rightleftharpoons_{\rm SLOCC}
|\Psi_{U(\tilde t)}\rangle$ (by calculation the corresponding normal
form for different $t$). This implies that $U(t)\not\rightleftharpoons
U(\tilde t)$ and simulation is impossible in both directions.

Note that one may also use \cite{noteVe} the results of Ref. \cite{Ve01b} 
to identify equivalence classes under SLOCC for multipartite 
unitary operations. There, it was shown that the problem reduces to establish 
whether two tensors are equivalent under LU.

\subsection{Many copy case:}

One may also apply Result 1 to the case where $N$ copies of the same
bipartite unitary operations $U$ should be performed simultaneously
und used to implement $M$ copies of some other unitary operation
$\tilde U$, i.e.  we investigate whether $U^{\otimes N}
\rightleftharpoons_{\rm SLOCC} \tilde U^{\otimes M}$ is possible.  For
example, one may want to know whether a single copy of a (strong
entangling) unitary operation, e.g. a CNOT gate, can be used to
implement $M$ (weakly entangling) operations, e.g. $\tilde
U=e^{-it\sigma_x^A\otimes\sigma_x^B}$ with $t\ll 1$. Given that
$n_{\Psi_{U_{\rm CNOT}}}=2$ and $n_{\Psi_{\tilde U^{\otimes M}}}=2^M$,
it follows from Result 1 that such a simulation is impossible, i.e $
U_{\rm CNOT} \not\rightharpoonup_{\rm SLOCC} \tilde U^{\otimes M}$,
even with an arbitrary small probability of success. This implies that
$U_{\rm CNOT} \not\rightleftharpoons_{\rm SLOCC} \tilde U^{\otimes
  M}$.  The reverse process, $\tilde U^{\otimes M}
\rightharpoonup_{\rm SLOCC} U_{\rm CNOT}$, is however possible.

This also implies that when demanding {\it exact} simulation,
bipartite unitary operations can even in the asymptotic case not be
reduced to a single one which serves as a representative for all kinds
of bipartite operations. One should ---as in the case of pure state
transformations in the asymptotic limit, where all kinds of bipartite
entanglement turn out to be equivalent to the one of the EPR-Bohm state---
allow for small imperfections. This still leaves open the possibility
that an equivalence relation under LOCC, allowing for some small
imperfections ---quantified e.g. by the fidelity for unitary gates as defined
in \cite{Vi01U,Ac01}---, such as
\be
U^{\otimes N} \approx\rightleftharpoons_{\rm LOCC} U_{\rm CNOT}^{\otimes M},
\ee
could exists. In this case, $M/N$ would be a measure for the non--locality
of $U$ and the CNOT operation could be used as an universal resource
to store without losses arbitrary bipartite unitary
operations. To proof or disproof such a relation would be of great interest.


\section{Generation of entangled states given U}\label{create}

Not only a classification based on (probabilistic) simulation of
operations might be of interest, sometimes a more practical approach
might be desirable. For example, one might want to know how powerful
a unitary operation is and which kind of entanglement such an
operation can create.  In this section, we will investigate such
questions and show which kind of entanglement a non--local unitary
operation $U$ can create probabilistically. The set--up we consider is
similar to the one of the previous section, i.e. we consider several
spatially separated parties, each possessing several $d$--level
systems. The initial state of the whole system is product. We may use,
before and after the application of $U$, an arbitrary sequence of
local operations and classical communication as well as arbitrary local
resources including auxillary systems. We are interested in the kind
of states which can be {\it probabilistically} created in this way,
that is we are interested in which kind of entanglement a given
unitary operation can produce under SLOCC. Note that we demand that
the state is created only with some nonzero probability of success.
Again, it turns out that this problem is closely related to the
problem of SLOCC conversion of pure states, which is expressed in the
following
  
{\bf Result 2}: $U$ can generate a state $|\Psi\rangle$ with
non-zero probability of success $\Longleftrightarrow$ $|\Psi_U\rangle$
can be converted to $|\Psi\rangle$ by means of SLOCC.

{\it Proof:}
$(\Rightarrow):$ We need to show that $|\Psi_U\rangle$ can be
converted to $|\Psi\rangle$ by means of SLOCC, given that $U$ can
generate the state $|\Psi\rangle$. This easily follows from the fact
that $|\Psi_U\rangle$ can be used to implement $U$ with certain
probability of success (see (\ref{iso2})), while $U$ can be used to
create $|\Psi\rangle$ out of a product state.\\
$(\Leftarrow):$ We have to show that $U$ can generate a state
$|\Psi\rangle$, given that $|\Psi\rangle$ can be obtained from
$|\Psi_U\rangle$ by means of SLOCC. Since $U$ can create the state
$|\Psi_U\rangle$ out of a product state (see (\ref{iso1})), which can
by assumption be transformed to the state $|\Psi\rangle$ by means of
SLOCC, the claim follows.

Let us illustrate Result 2 with help of some examples.

\subsection{Bipartite systems}

We consider a bipartite system of two $d$--level systems and an
arbitrary non--local unitary operation $U$. The Schmidt number
$n_{\Psi_U}$ of the corresponding state $|\Psi_U\rangle$ ---that is
the number of nonzero coefficients in the Schmidt decomposition---
completely determines which kind of bipartite entanglement can be
created by the unitary operation $U$. By means of SLOCC, all bipartite
states with lower or equal Schmidt number can be obtained from
$|\Psi_U\rangle$, while all other states cannot be created (see Sec.
\ref{SLOCCpure}). Thus it follows from Result 2 that $U$ can generate
all entangled pure states $|\Psi\rangle$ for which $n_\Psi \leq
n_{\Psi_U}$.  For example, the unitary operation
$U=e^{-it\sigma_x^A\otimes\sigma^B_x}$, $0<t<\pi/4$, can generate all
entangled pure states of Schmidt number 2, but cannot generate Schmidt
number 3 or 4 states. This follows from the fact that
$|\Psi_U\rangle=\cos(t)|\Phi_0\rangle_A|\Phi_0\rangle_B
-i\sin(t)|\Phi_1\rangle_A|\Phi_1\rangle_B$ has $n_{\Psi_U}=2$. 

Note that even if an operation is capable of creating Schmidt number 3 states, 
such states cannot be created directly. One first has to create a Schmidt number 4 
state which is then reduced to a Schmidt number 3 state by local measurements. 
This follows from the fact that the corresponding pure state of any unitary operation has either Schmidt number 1,2 or 4.

\subsection{Multipartite systems}

Consider as a second example a tripartite system of three qubits and a
non--local unitary operation of the form
\be
U=e^{-it\sigma_x^A\otimes\sigma^B_x\otimes\sigma^C_x}, \hbox{   } 0<t<\pi/4
\ee
We have that the corresponding state $|\Psi_U\rangle$ is given by 
\bea 
|\Psi_U\rangle&=&\cos(t)|\Phi_0\rangle_A|\Phi_0\rangle_B|\Phi_0\rangle_C\nonumber\\ 
&-&i\sin(t)|\Phi_1\rangle_A|\Phi_1\rangle_B|\Phi_1\rangle_C,
\eea
where the states $|\Phi_i\rangle$ are defined in Eq. (\ref{Bell}).
This is ---after a change of local basis $|\Phi_0\rangle\rightarrow
|0\rangle, |\Phi_1\rangle\rightarrow|1\rangle$ --- a state in
$(\C^2)^{\otimes 3}$ and thus effectively a state of three qubits.
Note that $|\Psi_U\rangle$ belongs to the GHZ--class and can thus not
be converted into the state $|W\rangle$ (see Sec.
\ref{SLOCCpuremulti}).  Using Result 2, this implies that $U$ cannot
create the state $|W\rangle$--- not even with a very small probability
of success.  However, $U$ can be used to create all states within the
GHZ-class.

Note that it also happens that three qubit unitary operation can
generate both kinds of of tripartite qubit entanglement. One such
example is the
unitary operation $U_W\equiv e^{-itH_W}$, $0<t<\pi/4$, where
$H_W\equiv|W\rangle\langle W|$ and the state $|W\rangle$ is defined in
Eq. (\ref{W}). Using that $H_W^2=H_W$, one readily observes that
$U_W=\eins+\gamma(t)H_W$, where
$\gamma(t)=\sum_{k=1}^\infty(-it)^k/k!$. It follows that
\bma\bea
U_W|001\rangle&=&|001\rangle+\gamma(t)/\sqrt{3}|W\rangle,\label{Ws}\\
U_W \frac{1}{\sqrt{2}}|0(0+1)1\rangle &=& \frac{1}{\sqrt{2}}(|001\rangle+|011\rangle+\frac{\gamma(t)}{\sqrt{3}}|W\rangle)\label{GHZs},
\eea\ema
where the state (\ref{Ws}) is a state in the W-class and the state
(\ref{GHZs}) is a state in the GHZ-class. Thus both states,
$|W\rangle$ and $|GHZ\rangle$, can be created probabilistically out of
a product state.

This can be understood as follows. The state $|\Psi_U\rangle$
corresponding to a tripartite unitary operation $U$ acting on three
qubits is in general a state acting on ${\cal H}=(\C^4)^{\otimes 3}$.
This implies that $U$ can create not only qubit--type entanglement
such as $|GHZ\rangle$ or $|W\rangle$, but also certain higher
dimensional entangled states, belonging in principle to different
classes under SLOCC. Recall that in tripartite four level systems, there
exist infinitely many inequivalent classes under invertible SLOCC
\cite{Du00W}. Although $|GHZ\rangle$ or $|W\rangle$ belong to
inequivalent classes under SLOCC, both of them may be created from a
higher dimensional state by means of non--invertible SLOCC, which
happens e.g. in the example discussed above. Similarly, maximally
entangled pure states shared between two parties, $A-B$, $A-C$ and
$B-C$ can be created from $|GHZ\rangle$ and $|W\rangle$ by means of
non--invertible SLOCC, although they belong to different equivalence
classes under invertible SLOCC.

\section{Equivalence classes for arbitrary operators}\label{General}

It is straightforward to generalize our results obtained for unitary
operations to the more general case of arbitrary operators. In
particular, since the isomorphism (\ref{iso}) also holds for
operators $O$, Results 1 and 2 also hold in this case and the proof
is exactly the same. Due to the fact that {\it any} pure state
$|\Psi_O\rangle \in \C^{\otimes d^2}$ corresponds to an
operator $O$ via Eq. (\ref{iso2}), we have that in contrast to what
happens for unitary operations $U$, exactly $d^2$ equivalence classes
under SLOCC for bipartite operators exist.

\section{Summary and conclusions}\label{Summary}

In this paper, we have provided a general framework to identify
equivalence classes of non--local unitary operations and arbitrary
operators under certain classes of local operations. For stochastic
local operations, assisted by classical communication, we provided
necessary and sufficient conditions for gate simulation in terms of
SLOCC conversion for pure states. This allowed us to obtain a
complete, hierarchic classification of bipartite unitary operations as
well as to obtain a number of results ---including a partial order---
for multipartite unitary operations. While for bipartite operations
always a finite number of inequivalent classes under SLOCC exists, we
showed that for multipartite operations one obtains
infinitely many classes.  The important case of bipartite unitary
operations acting on qubits was studied in detail, and we identified
three different kinds of bipartite unitary operations under SLOCC,
represented by product operations, the CNOT operation and the SWAP
operation respectively.

We also showed which kind of entanglement a unitary operation can
create. Again, this was done by obtaining a connection to the problem
of state conversion under SLOCC. We provided a complete solution in
the bipartite case and discussed some implications for the
multipartite setting. 

The problem of deterministic simulation of non--local unitary
operations under LU and LOCC was only discussed briefly and some
future work is desirable. In particular, it would be
interesting to obtain normal forms for high dimensional bipartite
unitary operations under LU, i.e. to identify the corresponding
equivalence classes, and to identify equivalence classes under
LOCC in the single and multi--copy case.

\section*{Acknowledgements}
We would like to thank G. Vidal for many useful discussions and comments.
This work was supported by European Community under project EQUIP (contract 
IST-1999-11053) and through grant HPMF-CT-2001-01209 (W.D., Marie Curie fellowship),  the ESF and 
the Institute for Quantum Information GmbH .



\begin{references}
\bibitem{noteCNOT}
The CNOT operation is defined by the following mapping of states, written in the
standard basis: $|i\rangle_A|j\rangle_B \rightarrow |i\rangle_A|i\oplus
j\rangle_B$, where $\oplus$ denotes addition modulo 2. The SWAP
operation is given by the mapping  $|i\rangle_A|j\rangle_B \rightarrow |j\rangle_A|i\rangle_B$
\bibitem{noteunique}
The uniqueness of this decomposition can easily derived using the results of Ref. \cite{Kr00}.
\end{references}
\end{document}